\documentclass[prl,twocolumn]{revtex4}
\usepackage{graphicx}
\begin{document}

\title{Onset of dissipation in ballistic atomic wires}

\author{Nicol\'{a}s Agra\"{\i}t}%
        \email{nicolas.agrait@uam.es}
\author{Carlos Untiedt}
        \altaffiliation{Present address: Kamerlingh Onnes Laboratorium,
        Universiteit Leiden, Postbus 9504, NL-2300 RA Leiden, The Netherlands}

\author{Gabino Rubio-Bollinger}
\author{Sebasti\'{a}n Vieira}
\affiliation{Laboratorio de Bajas Temperaturas,\\
Departamento de F\'{\i}sica de la Materia Condensada C-III,\\ and
Instituto Universitario de Ciencia de Materiales ``Nicol\'{a}s
Cabrera'', \\ Universidad Aut\'{o}noma de Madrid, E-28049 Madrid,
Spain}

\date{\today}
\begin{abstract}
Electronic transport at finite voltages in free-standing gold
atomic chains of up to 7 atoms in length is studied at low
temperatures using a scanning tunneling microscope (STM). The
conductance vs voltage curves show that transport in these
single-mode ballistic atomic wires is non-dissipative up to a
finite voltage threshold of the order of  several mV. The onset of
dissipation and resistance within the wire corresponds to the
excitation of the atomic vibrations by the electrons traversing the
wire and is very sensitive to strain.
\end{abstract}


\maketitle

The trend of miniaturization in electronics will soon lead to
devices of  nanometer-scale in which quantum effects become
relevant. The ultimate quantum conductor is a perfect
one-dimensional wire, such as an atomic chain \cite{1, 2} or
semiconducting heterostructure \cite{3}. In these wires electrons
are ballistic since there are no defects to inhibit resistance-free
currents \cite{3}. The limiting factor in the current-carrying
capacity of a wire is dissipation, which results in heating. Two
mechanisms contribute to the resistance of a metallic wire: elastic
scattering with defects and impurities and inelastic scattering
with the lattice vibrations \cite{4}. In the absence of scattering,
electrons can propagate freely and transport is said to be
ballistic. This situation is possible in the nanoscale where the
mean free path of electrons can be much longer than the length of
the device.

The two-terminal zero-bias resistance of a single mode ballistic
wire is the resistance quantum $h/2e^{2}$. This resistance is
entirely associated with the connections of the wire to the
electrodes \cite{5}, being the intrinsic resistance of the wire
zero, as recently demonstrated in quantum wires fabricated from
GaAs/AlGaAs heterostructures \cite{3}, and in agreement with
Landauer framework \cite{6, 7}. Within this framework, the applied
voltage serves to unbalance the chemical potentials for propagating
electrons in each direction and drops entirely at the contacts and
not within the wire. The Joule dissipation associated to this
resistance is assumed to take place far away from the contact (at
an inelastic relaxation length) where electrons and holes relax to
the Fermi level of the electrodes.  This picture is correct for
bias voltages close to zero, which implies vanishingly small
currents (note that the resistance free currents in the experiment
of ref. \onlinecite{3} were smaller than 1 nA).

In this letter we study transport at finite voltages and the
mechanism of dissipation in ballistic wires. Our experiments are
performed in freely suspended gold atomic wires of up to 7 atoms in
length, fabricated using a low-temperature scanning tunneling
microscope \cite{1, 2}. Very recently the forces and conductance
have been measured simultaneously \cite{8} during the process of
chain formation giving insight into the formation mechanisms. The
mechanical and electronic properties of these metallic
nanostructures are of great interest not only from the point of
view of the applications but also from a fundamental point of view,
since the system has few atoms and should be amenable to detailed
modelling.

Atomic wires of gold are fabricated at low temperatures (4.2 K),
using an STM. Fabrication of an atomic wire is as follows \cite{1}:
a metallic contact is formed between a tip and a substrate of an
STM, both made out of Au, with 99.99 \% purity. Elongation of the
contact results in a decrease in its size. In the final stage a
one-atom contact is formed, and as we continue elongating there is
a certain probability that an atomic chain forms \cite{1}. The
length of the wire can be estimated from the length of the last
plateau before rupture, whose conductance is approximately one in
units of $2e^2/ h$. The atomic wires are very stable at low
temperature and the measured curves are completely reproducible as
long as the tip position is not changed. Typically an experiment on
a given atomic wire takes about half an hour, and ends when the
wire breaks as a result of further elongation. The differential
conductance $G$ of the wires as a function of voltage is measured
using a lock-in technique, with a small modulation of 1 mV. The
derivative of the differential conductance $dG/dV$ is calculated
numerically. The energy resolution of our measurement is thermally
limited to 2 meV. Sharper peaks would be expected for lower
temperatures, but a finite with could be due to the finite length
of the wire.

\begin{figure*}
\includegraphics{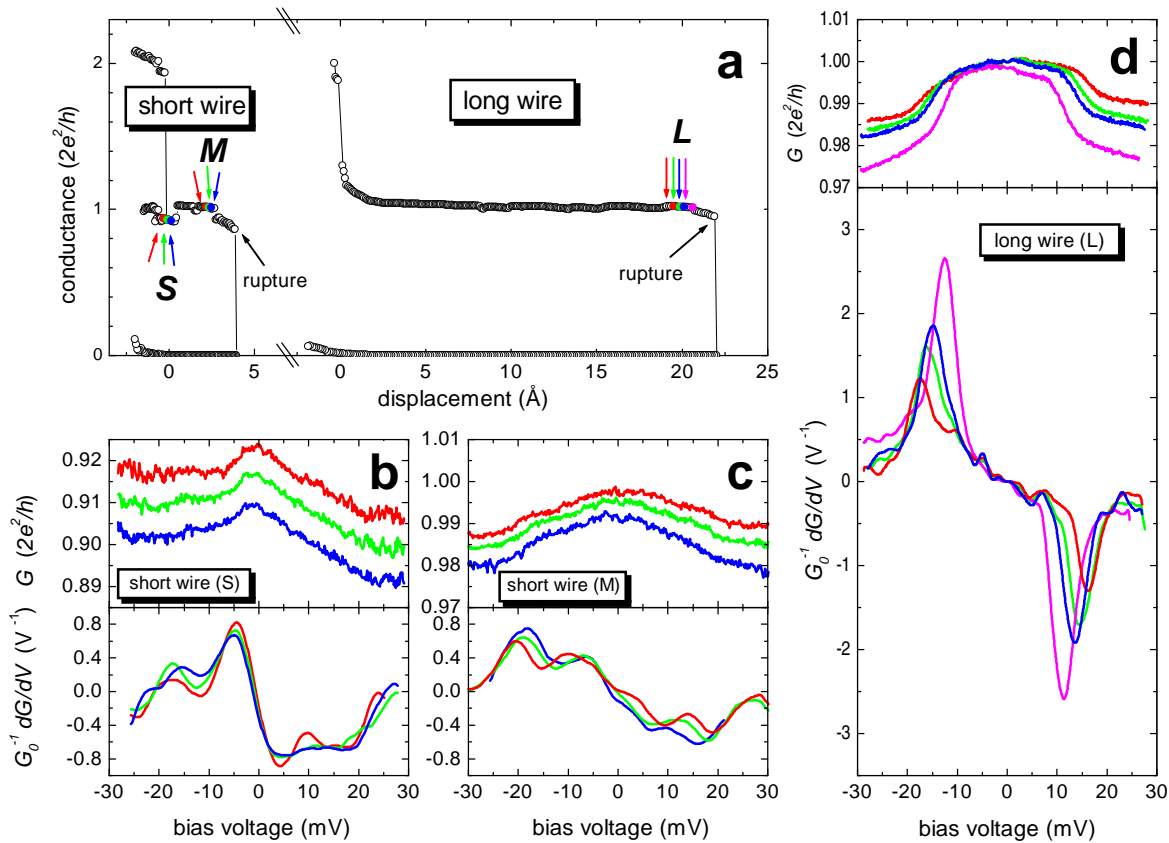}
\caption{\label{fig1} (a) Short and long atomic wire, $\sim 4$ \AA\ long and $\sim 22$
\AA\ long, respectively. Panels (b), (c) and (d) show the differential
conductance and its derivative at points S, M, and L, respectively,
marked by the arrows. The various curves in (b), (c), and (d) were
acquired at intervals of 0.3, 0.3 and 0.5 \AA, respectively. Note
that the vertical scales are identical in these panels. }
\end{figure*}

Experimentally, it is observed \cite{1} that the zero bias
conductance of Au wires of up to 7 atoms in length is close to
$G_{0}=2e^{2}/h$, the quantum conductance unit, independently of
their length, reflecting the fact that gold wires have a single
almost completely open quantum channel, as confirmed by theoretical
calculations \cite{9,10,11}. This agrees with Picciotto {\em et
al.} \cite{3}, showing that the resistance is at the contacts, not
within the chain. These atomic wires do not show an ohmic
behaviour: the conductance is voltage dependent. We measure the
differential conductance $G = dI / dV$, of atomic wires of
different lengths (from 1 to 7 atoms). Typical differential
conductance curves for short and long atomic wires are shown in
Fig.~\ref{fig1}(b)-(d). The differential conductance $G$ shows a
hump at zero bias, dropping about 1\% in the range (20 mV). The
asymmetry in $G$ has been shown to be due to elastic scattering
which results in interference effects \cite{12,13}.

\begin{figure}
\includegraphics[width=8cm]{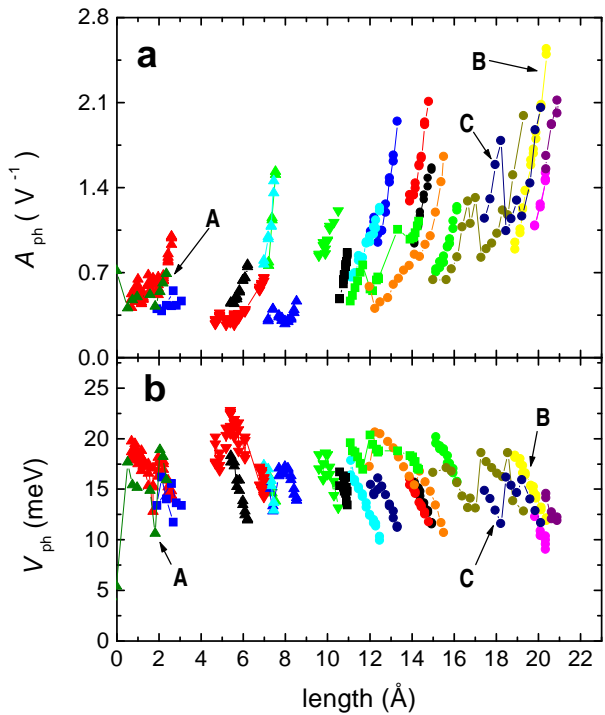}
\caption{\label{fig2} (a) Magnitude $A_{\text{ph}}$  and
(b) position $V_{\text{ph}}$ of the phonon peak in the PC spectrum
as a function of length chain. Here we present 22 different chains
(out of more than 100 studied). Each chain is represented by a
different symbol. The chains labeled A and B are the short and long
wires, respectively, of Fig.~\ref{fig1}; and C is the wire in
Fig.~\ref{fig3}. The length of the chain is estimated from the
length of the last plateau.}
\end{figure}

These symmetric drops in the conductance are characteristic of
inelastic scattering of electrons, and the range of voltages is
typical of phonons. At cryogenic temperatures, the differential
conductance of larger ballistic metallic contacts (100--1000 atoms
in cross-section) is also voltage dependent, and has been used to
obtain energy resolved information on the interaction of electrons
with phonons and other elementary excitations. This
well-established technique is known as point-contact (PC)
spectroscopy \cite{14, 15}. When a voltage drop $V$ is applied to a
metallic contact, whose radius a is smaller than the electron mean
free path,  electrons are injected into the higher voltage
electrode with excess energies of up to $eV$. These electrons have
a finite probability of exciting a vibrational mode of the lattice,
{\em i.e.,} emitting a phonon. At these low temperatures the
probability of phonon absorption is very low because the
equilibrium phonon population is almost zero. The resulting $dG/dV$
{\em vs} voltage curves, or PC spectra, are proportional to the
phonon density of states (DOS) times the electron-phonon coupling
strength. The spectra scale with the zero bias conductance as
$\propto G_{0}^{3/2}$, reflecting the fact that only electrons that
are scattered close to the contact, {\em i.e.} within a distance to
the contact of the order of the contact radius, have a finite
probability of being backscattered, and consequently of being
detected as a reduction in conductance. The spectra for Au have
peaks at 10 mV and 18 mV, corresponding to the maxima in the
transverse  and longitudinal phonon DOS at 10 meV and 18 meV,
respectively. The transverse peak being stronger than the
longitudinal peak.

In contrast to the PC spectra of larger ballistic contacts, the
spectra of atomic wires are very sensitive to the atomic
configuration at the contact. Fig.~\ref{fig1}(b) and (c) show the
differential conductance and spectra at two different situations
during the elongation of a short atomic wire, positions labeled S
and M in Fig.~\ref{fig1}(a), respectively. At S the atomic wire is
likely to be just one atom long, while at M the contact has been
further stretched about 3
\AA\  that is, the wire is about two atoms long. The evident
broadening of the conductance hump is typically observed for atomic
wires longer than one atom. The signal is about three times larger
than that given by the semiclassical theory of PC spectroscopy
\cite{15}. Fig.~\ref{fig1}(d), corresponds to the long wire (about
7 atoms long) in Fig.~\ref{fig1}(a). For this long wire the
conductance drop takes place quite sharply, which results in a
sharp peak in the spectra. This dependence of the spectra on the
length of the wire indicates that most of the measured signal comes
from processes occurring within the wire itself.

\begin{figure}[b]
\includegraphics[width=8cm]{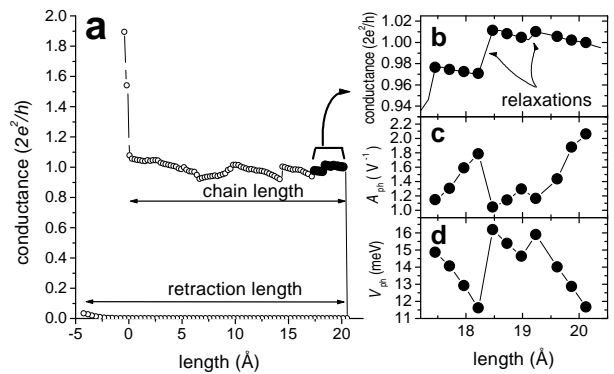}
\caption{\label{fig3} (a) Evolution of the conductance for a 20 \AA-long atomic
wire. (b) Zoom of the last part of this evolution showing abrupt
jumps in the conductance corresponding to force relaxations due
atomic rearrangements between the elastic stages.  (c) and (d) show
the magnitude $A_{\text{ph}}$ and position $V_{\text{ph}}$,
respectively, of the phonon peak in the PC spectra.}
\end{figure}

The observed spectra for the longer atomic wires are a signature of
their one-dimensionality. Due to momentum conservation, in a
one-dimensional conductor electrons can only excite vibrations of
the atomic chain whose wavenumber is twice the Fermi wavenumber
$k_{F}$. In this process an electron dissipates its energy losing
one quantum of vibrational energy $\hbar \omega_{2k_{F}}$  and
being backscattered. Here, $\omega 2k_{F}$ is the frequency of the
excited vibrational mode. In a ballistic conductor, this phonon
emission process will only be possible for voltages $V$ larger than
$V_{\text{ph}} =\hbar \omega_{2k_{F}} /e$, since the chemical
potentials for right-going electrons and for left-going electrons
are unbalanced by $eV$. Consequently, the onset of dissipation is
marked by a sudden decrease in the conductance, since the
backscattered electrons do not contribute to the current. The
position of the peak in the spectra  $V_{\text{ph}}$  gives the
frequency of the $2k_{F}$-phonon and its height $A_{\text{ph}}$,
which is related to the conductance drop, is proportional to the
probability of the phonon emission process. Ideally, for infinite
wires, the conductance drop would be sharp. In our measurements we
have an energy resolution thermally limited to about 2 meV and the
finiteness of the wire would also result in a finite width. The
magnitude of the conductance drop (about 1\% for a chain of 20
\AA\ in length) is consistent with an inelastic mean free path of
about 2000 \AA\ in an infinite wire, which is reasonable for a
metal at low temperatures.

Experimentally, it is observed that the position and amplitude of
the peak in the spectra of a given atomic wire is very sensitive to
its state of strain. As shown in Fig.~\ref{fig1}(d), stretching the
wire results in a decrease of the frequency of the phonon and an
increase in the emission probability. This phonon softening
indicates that the elastic constant of an atomic chain decreases as
it is stretched. The increase in the emission probability indicates
an enhancement of the electron-phonon interaction. In
Fig.~\ref{fig2}, we have plotted the position of the peak in the
spectra and its amplitude for many different atomic wires. The
emission probability increases with the length of the wire, and the
variations due to stretching are much larger than for short wires.
This is an indication that most of the signal comes from the wire
itself.

An atomic wire can be stretched elastically only a limited distance
($\sim$ 1 \AA) before breaking or changing to a new configuration.
These configurational changes result in stress relaxations
correlated to abrupt conductance jumps \cite{8}, whose magnitude is
much smaller than a conductance quantum $2e^{2}/h$. Some of the
stress relaxations correspond to the incorporation of an extra atom
into the atomic wire, while other relaxations come from atomic
rearrangements occurring in the electrode region close to the wire
\cite{8}. In Fig.~\ref{fig3}(b), (c), and (d), we show the
evolution of the conductance, and the amplitude and position (in
energy) of the phonon peak as the wire is elongated along 3 \AA\
before chain rupture. The final length of the wire was 21 \AA\ [see
Fig.~\ref{fig3}(a)]. Note that jumps in the amplitude and position
of the phonon peak in the sprectra are correlated to conductance
jumps and consequently to stress relaxations. The observed phonon
softening with elastic deformation is consistent with the results
of the ab initio calculations of S\'{a}nchez-Portal {\em et al.}
\cite{16} for linear and zig-zag atomic wires. Our results are
compatible with both types of wires. Detailed theoretical
calculations would be necessary for a quantitative understanding of
the electron-phonon interaction and its enhancement with elastic
strain in metallic atomic wires.

As we have seen dissipation in one-dimensional atomic wires occurs
through the process of phonon emission. In contrast to what happens
in a three dimensional geometry, the hot electrons can excite a
single vibrational mode of the atoms of the wire, and this requires
electrons with sufficient energy. As a consequence, for voltages
below  $V_{\text{ph}}$, no dissipation occurs within the atomic
wire. The maximum dissipationless current that an atomic wire can
carry is then $G_{0} V_{\text{ph}}\approx 1 \mu$A, which is an
enormous current density $\sim 10^{7}$ A/mm$^2$. For voltages above
$V_{\text{ph}}$ , the wire progressively heats up. For a given
voltage a balance is established among the energy released by the
electrons, the energy absorbed by the electrons (phonon absorbtion)
and the energy lost by thermal conduction to the electrodes. In our
experiments, atomic chains could sustain voltages of up to 500 mV
(note that in the same conditions an atomic contact could stand
2000 mV).

In summary, we have studied the voltage dependence of the
conductance of metallic atomic wires at low temperatures, showing
that inelastic scattering of electrons sets in at a finite voltage
due to the excitation of the vibrations of the ions of the atomic
chain. This well-defined threshold for dissipation is
characteristic of the electron-phonon interaction in
one-dimensional systems. We observe that the mechanical tensioning
of the atomic chain results in bond-softening, which reflects in a
decrease of the phonon frequency, and a dramatic enhancement of the
electron-phonon interaction.

\begin{acknowledgments}
This work has been supported by CICyT grant PB97-0068.
\end{acknowledgments}


\end{document}